\documentclass[twocolumn]{aastex63}
\usepackage[utf8]{inputenc}

\submitjournal{BAAS}

\shorttitle{Socio-demographic study of the exoplanet direct imaging community}
\shortauthors{Leboulleux, Choquet, Huby, Singh, Cantalloube}

\graphicspath{{./}{figures/}}

\begin{document}

\title{A socio-demographic study of the exoplanet direct imaging community 
}

\correspondingauthor{Lucie Leboulleux}
\email{lucie.leboulleux@obspm.fr}

\author{Lucie Leboulleux}
\affiliation{LESIA, Observatoire de Paris, Universit\'e PSL, CNRS, Université de Paris, Sorbonne Universit\'e, \\ 5 place Jules Janssen, 92195 Meudon, France}

\author{Élodie Choquet}
\affiliation{Aix Marseille Univ, CNRS, CNES, LAM, Marseille, France}

\author{Elsa Huby}
\affiliation{LESIA, Observatoire de Paris, Universit\'e PSL, CNRS, Université de Paris, Sorbonne Universit\'e, \\ 5 place Jules Janssen, 92195 Meudon, France}

\author{Garima Singh}
\affiliation{LESIA, Observatoire de Paris, Universit\'e PSL, CNRS, Université de Paris, Sorbonne Universit\'e, \\ 5 place Jules Janssen, 92195 Meudon, France}

\author{Faustine Cantalloube}
\affiliation{Max Planck Institute for Astronomy, Königstuhl 17, 69117 Heidelberg, Germany}

\begin{abstract}

Astronomy and science are fields in which specific groups remain under-represented despite multiple studies that investigate this issue and propose solutions. In this article, we analyze the demographics and social behavior of the exoplanet direct imaging community. Our focus is on identifying possible under-representation among this group, and quantifying inappropriate social behaviors. During the Spirit of Lyot conference 2019 (Tokyo, Japan), we conducted a survey that gathered a participation rate of 53\%. We analyzed the data collected under the prisms of gender balance and seniority representation. The proportions of women and of non-binary persons reveal a more diverse community in comparison to the other scientific groups (e.g. the IAU members), but still far from a balanced representation of all genders. Early-career scientists appear to have a lower visibility in the field than permanent researchers, with PhD students being under-represented at international conferences, and postdocs being excluded from conference Science Organizing Committees. Regarding social relations, the results are alarming, in particular when it comes to self-censoring of women or to unprofessional behavior, which was experienced by 54\% of this community (gender-biased behavior: 29\%; oral interruption: 33\%; inappropriate behavior: 33\%), and in particular by women. We recommend the community to become pro-active to build a safe environment and to continue its inclusion efforts. One aspect could be to systematically include socio-demographic surveys in conference registration forms to monitor the evolution of the community, in particular at larger scales. To do so, the survey questions available on GitHub.

\end{abstract}


\section{Introduction} \label{sec:intro}

Science, Technology, Engineering and Mathematics (STEM) are fields traditionally affected by a low workforce diversity, despite inclusion being repeatedly pointed out as necessary to increase the performance and the quality of a workplace \citep{Slater2008}. 
In particular, the gender gap has been studied from various angles: access to permanent positions \citep{BerneHilaire2020}, responsibilities \citep{Rathbun2015, Rathbun2017, Piccialli2019}, proposal acceptance rate \citep{Reid2014, Lonsdale2016, Patat2016, Spekkens2018}, conference attendance \citep{Davenport2014, Prichard2019}, citations  \citep{Caplar2017}, etc. -- and on various scales: in general science and technology  \citep{Casadevall2014}, in astronomy \citep{Cesarsky2010, Norman2019}, at the scale of a country \citep{Spekkens2018, BerneHilaire2020}, of an instrument \citep{Reid2014, Lonsdale2016}, of an institute \citep{DeRosa2019, Hibon2018}, of a sub-field of astronomy \citep{DOrgeville2014}, etc. Despite these multiple studies, gender imbalance is still significant and an ongoing issue. 
In order to stimulate changes of behavior and improve the situation, under-representation and biased behaviors need to be recognized as an issue, outspoken, and monitored within the STEM community rather than minimized or disregarded comparatively to scientific questions.
To do so, more studies are needed to monitor the evolution of the demographics and of the behaviors, and to extend the awareness to other minorities under-represented in science.

In astronomy, several recent studies focused on factors impacting the visibility of women and having a direct effect on their professional evolution and/or recognition. In particular, \citet{DOrgeville2014} provided a detailed overview of the reasons causing women to leave astronomy and more specifically the field of adaptive optics. The authors pointed out numerous factors including social misconduct towards women and the impostor syndrome of which women are in majority subjects in this field. They also formulated a number of propositions to counterbalance this effect in the future. Other studies have focused on gender-based biases within selection committees that affect the visibility of women in astronomy, such as conference speakers selection, for instance at the American Astronomical Society (AAS) meeting 223 \citep{Davenport2014} or at the 2014 to 2016 American Geophysical Union (AGU) Fall Meetings \citep{Ford2018}, or observing program selection, for example the Hubble Space Telescope \citep{Reid2014} and the ESO \citep{Patat2016} time allocations. These biases observed in astronomy are similar to what is seen in other fields in Science: in microbiology for instance, a study by \citet{Casadevall2014} also suggests that the gender balance within Science Organizing Committees (SOC) and conference conveners directly impacts the distribution of talks per gender at conferences, which has also been observed by \citet{Nittrouer2018} in six disciplines (biology, bio-engineering, political science, history, psychology, and sociology).

Other studies specifically addressed the impact of seniority on the gender balance in astronomy and pointed out important dependencies between these two aspects on demographic and behavior questions. For instance, \citet{Cesarsky2010} showed that the gender ratio evolves with  age or  career status, suggesting that some gender-based results are degenerate with the career level of the probed population. A few studies focused on disentangling the career stage with the gender. For instance, \citet{Spekkens2018} showed that during the Canadian time allocation process, gender is the only significant discrimination parameter.

These numerous studies illustrate the importance of staying alert about inequalities towards women and minorities in Science in general and in Astronomy in particular. Capturing the demographics of a community is the first step to: 1) identify under-representation issues, 2) develop solutions to improve the inclusion of all groups and, 3) set a reference point to monitor the (hopefully positive) evolution of this community.
For these reasons and to complement similar studies in other sub-fields of astronomy, we probed the attendees of the Spirit of Lyot 2019 conference held in Tokyo, Japan, and obtained the first demographic snapshot of the Exoplanet Direct Imaging community. 

The Spirit of Lyot (SOL) conference is a major international conference gathering the community studying extrasolar systems with high-contrast imaging instruments. Its main motivation is to bring together researchers with different expertise ranging from observation to instrumental research and development working towards the same scientific objective: the detection and characterization of exoplanetary systems.
This conference brings together a large fraction of this community on a 4-year basis. The fourth edition of the SOL conference was held in Tokyo in October 2019 where around 200 researchers participated. Here, we report the outcome of the survey that was shared with all the attendees of the conference, thus providing a large and representative sample of the field of Direct Imaging of Exoplanets.

In section \ref{sec:Method}, we describe the survey proposed at the SOL conference and the methodology used to analyze the data. Section \ref{sec:GeneralDemographics} presents the general demographic overview  of the participants. Sections \ref{sec:Gender} and \ref{sec:Status} describe and analyze the results of the survey as a function of  the gender and of the career position of the participants, respectively. The conclusion section  summarizes the outcomes of our study, identifies its main limitations and proposes solutions to overcome them.


\section{Methodology}
\label{sec:Method} 

The results presented in this report were obtained from a survey conducted in October 2019 at the Spirit of Lyot conference in Tokyo, Japan. The survey was initiated during an unofficial splinter meeting which included around 12 female and non binary participants interested in gender studies. Discussions during this meeting enabled to identify broad categories as well as specific points interesting to be probed. 

The complete list of questions asked in the survey is presented in appendix~\ref{sec:Appendix}. An improved version is available on open source on GitHub: \href{https://github.com/lleboulleux/socio-demographic-community-survey-in-STEM}{https://github.com/lleboulleux/socio-demographic-community-survey-in-STEM}. Four main topics were addressed: the general demographics in the field, the visibility and the ability of the participants to self-promote during the SOL conference, their visibility and their recognition in the field in general, and the occurrence of unprofessional behaviors in this community. While most of the questions were objective and factual queries, a few others (questions 12 to 15, 17, 19, and 20) focused on the subjectivity or on the perception of the participants. We should point that none of the authors has a background in social science and the formulation and topics addressed by the questions may benefit from researchers in social science to guaranty additional neutrality. We encourage the community and in particular demographic studies experts to either commit edits to the form in the Github repository and/or directly contact the authors to improve the survey. We further discuss the implications in Sec. \ref{sec:Gender} and \ref{sec:Conclusions}, respectively.

The survey was opened on October 24th 2019 on the fourth day of conference, and sent to all of the  conference participants. It was advertised by the SOC between conference sessions and by the Local Organizing Committe (LOC) on the conference email list. The survey was open for three weeks and was closed on November 14th. A reminder email was sent to the conference participants on November 4th, at midpoint of survey open period. All the answers were collected on a voluntary basis, which may bias the accuracy of the results as some groups may be more responsive than others \citep[e.g][]{DOrgeville2014}. These limitations are discussed specifically in sections  \ref{sec:Gender} and \ref{sec:Status}.

In the following sections, we analyze the answer ratios of each question, first for the participants of the survey as a whole, then for specific groups among the participants. We computed uncertainties on these ratios at $66\%$ confidence level following the study by \citet{DOrgeville2014}, with uncertainties given by $\sigma = \sqrt{\frac{M(N-M)}{N}}$, where $N$ is the total number of people in the study group and $M$ the number of people of the specific category analyzed in this study group. For instance, $N$ could be the number of women while $M$ could be the number of people having answered "yes" to a specific question in the women group. 


\section{General demographics}
\label{sec:GeneralDemographics} 

We collected precisely 100 answers to the survey out of a total of 190 conference participants. This gives a participation rate  of 53\%, a value high enough to provide statistical results representing the SOL 2019 conference participants.

\begin{figure*}
   \begin{center}
   \begin{tabular}{cc}
   \includegraphics[height=4.5cm]{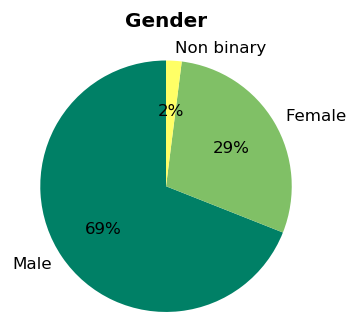} & \includegraphics[height=4.5cm]{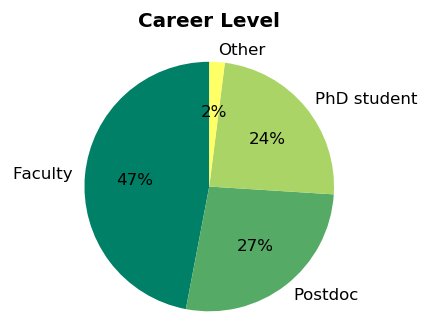} \\
   \includegraphics[height=5.25cm]{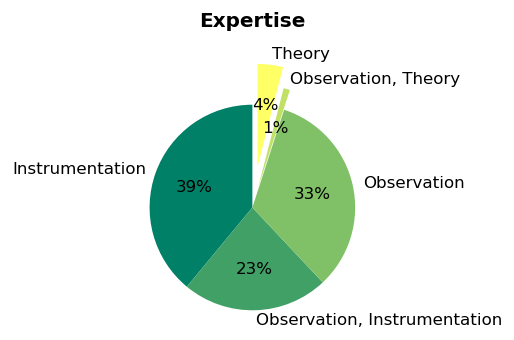} & \includegraphics[height=5.25cm]{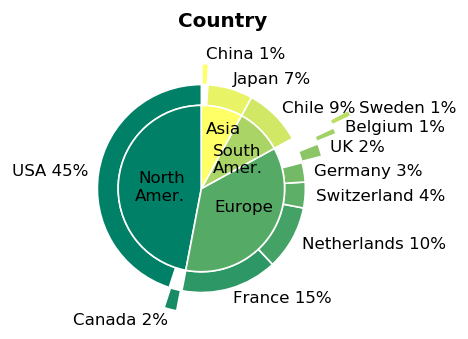}
   \end{tabular}
   \end{center}
   \caption 
   { \label{fig:General} 
Overall demographics of the participants of the Spirit of Lyot 2019 conference: these graphics show the ratios of the participant to the survey as a function of their gender, career level, field of expertise, and affiliation country.}
   \end{figure*}

In Fig.~\ref{fig:General} we present the main demographic characteristics obtained from these 100 answers.
%
The gender balance in the survey participants is as follows: 69\% defined themselves as men, 29\% as women, and 2\% as non-binary people. These proportions confirm the under-representation of women in this field and are comparable to the ones in the other fields presented in the introduction (e.g. 27\% of women individuals in the field of Adaptive Optics, \cite{DOrgeville2014}). Compared to the whole field of Astronomy, this proportion of women in the field of Exoplanet Direct Imaging is lower than the one in AAS  \citep[34\%,][]{Davenport2014}, but higher than the fraction of women in the IAU (18\%, data from January 2018) and equivalent to the one attending the AGU Fall Meetings \citep[28\%,][]{Ford2018}. The presence of non-binary people is also encouraging, but can hardly be compared to any reference percentage since the studies including them are unfortunately too rare and should be developed, which \cite{Rasmussen2019} and \cite{Strauss2020} propose recommendations for. Although encouraging, the representation of women and non binary people at the SOL conference shows that significant efforts need to be performed to reach an acceptable gender balance in this community.

In terms of career level distribution, 24\% of the participants were PhD students, 27\% post-doctorate researchers, 47\% faculty researchers, and 2 people did not answer this question. The participation to the conference was thus at 51\% for young scientists and 47\% for permanent researchers. If we assume an average ratio of two non-permanent positions per permanent position in the community as deduced from observations in different institutes and accounting for geographical disparities (higher ratio in the USA than in European countries), we notice a significant under-representation of young-career scientists at the 2019 SOL conference. In comparison, the representation of PhD students at the AGU Fall meetings was 29\% between 2014 and 2016, a ratio 5 points higher than at the Spirit of Lyot 2019 conference \citep{Ford2018}. This indicates a community which favors self-promotion (faculties are in charge of managing travel budgets) over the promotion of young scientists in the team. This is perceived as a negative point for this community given that PhD students and postdocs strongly rely on international conferences to progress in their career and gain visibility.

The expertise of the conference attendees was fairly spread over three main domains: instrumentation (39\%), observations (33\%), and a combined observational and instrumental expertise (23\%). This reflects the rational of the Spirit of Lyot conference series, which specifically aims at bringing together these fields of expertise related to the direct imaging of exoplanetary systems, a field where astrophysical results depend on complex instruments. In addition, a low but noticeable fraction of the attendees claimed expertise in theory (5\%), indicating that the field also requires theoretical work to interpret their observations. Monitoring the progress of the fraction of theoreticians over time may indicate an interesting evolution of the field toward more  fundamental research about the formation and evolution of exoplanets and circumstellar disks.

Finally, the participants of the 2019 SOL conference came in majority from the USA (45\%) and from Europe (36\%, mainly France and the Netherlands at 15\% and 10\%, respectively). The other continents were much less represented, with only 9\% of the participants coming from South America (exclusively Chile), 8\% from Asia (mainly Japan, 7\%), and no participant from Africa or Oceania. This is a negative sign in terms of the geographic representation in this field, in particular considering that several major instruments used by this community are installed in South America and the conference was held in Asia. Among the possible reasons for these geographic disparities are  financial reasons (different travel budgets), the importance of the field in each country (also linked to the financial reasons through hiring resources), and interest in the conference within each country.

In Tables \ref{tab:Country_Gender} and \ref{tab:Country_Career}, we show the local distributions of gender and career level, respectively, among the participants as a function of their affiliation country. We excluded the countries with only one answer to the survey from this analysis (Belgium, China, Sweden). Although some ratios are affected by small sample statistics and should be used carefully, significant  differences appear between some countries in terms of gender career category representation. In particular, the three most represented  countries, with more than 10\% of the participants, were the USA, France, and the Netherlands. We see important differences between them: French participants were more balanced in gender than the other two countries with 40\% female scientists, but more strongly favored the participation of the permanent researchers over young scientists (40\%). Conversely, Dutch participants had a large majority of male participants (only 20\% of women), but promoted their students and postdocs more strongly (80\% of young scientists) over the faculties. The US participants were more balanced on these categories, and included the only non binary scientists who answered the survey, and thus showed a more diverse environment. However, their female and young-career scientists remain under-represented, with only 24\% of women, and 47\% of PhD students and postdocs.

\begin{table}
    \caption{Gender distribution among the participants for each country with more than one respondent to the survey.}
    \label{tab:Country_Gender}
    \centering
    \begin{tabular}{lrrr}
        \hline \hline
        Country & Female & Male &   Non binary \\
                &  (\%) & (\%)  & (\%)\\
        \hline 
         USA    & $24\pm6$ & $71\pm7$  & $4\pm3$\\
         France & $40\pm13$ & $60\pm13$  & $0\pm0$\\
         Netherlands & $20\pm13$ & $80\pm13$  & $0\pm0$\\
         Chile & $33\pm16$ & $67\pm16$  & $0\pm0$\\
         Japan & $43\pm19$ & $57\pm19$  & $0\pm0$\\
         Switzerland & $50\pm25$ & $50\pm25$  & $0\pm0$\\
         Germany & $33\pm27$ & $67\pm27$  & $0\pm0$\\
         Canada & $0\pm0$ & $100\pm0$  & $0\pm0$\\
         United Kingdom & $50\pm35$ & $50\pm35$  & $0\pm0$\\
    \end{tabular}
\end{table}

\begin{table}
    \caption{Career level distribution among the participants for each country with more than one respondent to the survey.}
    \label{tab:Country_Career}
    \centering
    \begin{tabular}{lrrr}
        \hline \hline
        Country & Faculty & Postdoc &   PhD Student  \\
                &  (\%)         & (\%)       & (\%)\\
        \hline 
         USA    & $49\pm7$ & $29\pm7$  & $18\pm6$\\
         France & $60\pm13$ & $20\pm10$  & $20\pm10$\\
         Netherlands & $20\pm13$ & $30\pm14$  & $50\pm16$\\
         Chile & $67\pm16$ & $11\pm10$  & $22\pm14$\\
         Japan & $57\pm19$ & $14\pm13$  & $29\pm17$\\
         Switzerland & $0\pm0$ & $100\pm0$  & $0\pm0$\\
         Germany & $33\pm27$ & $33\pm27$  & $33\pm27$\\
         Canada & $50\pm35$ & $0\pm0$  & $50\pm35$\\
         United Kingdom & $50\pm35$ & $0\pm0$  & $50\pm35$\\
    \end{tabular}

\end{table}


\section{Gender-based analysis}
\label{sec:Gender} 

\subsection{Involvement of the participants}
\label{sec:GenderImplication}

\begin{figure*}
   \begin{center}
   \begin{tabular}{cc}
   \includegraphics[height=5cm]{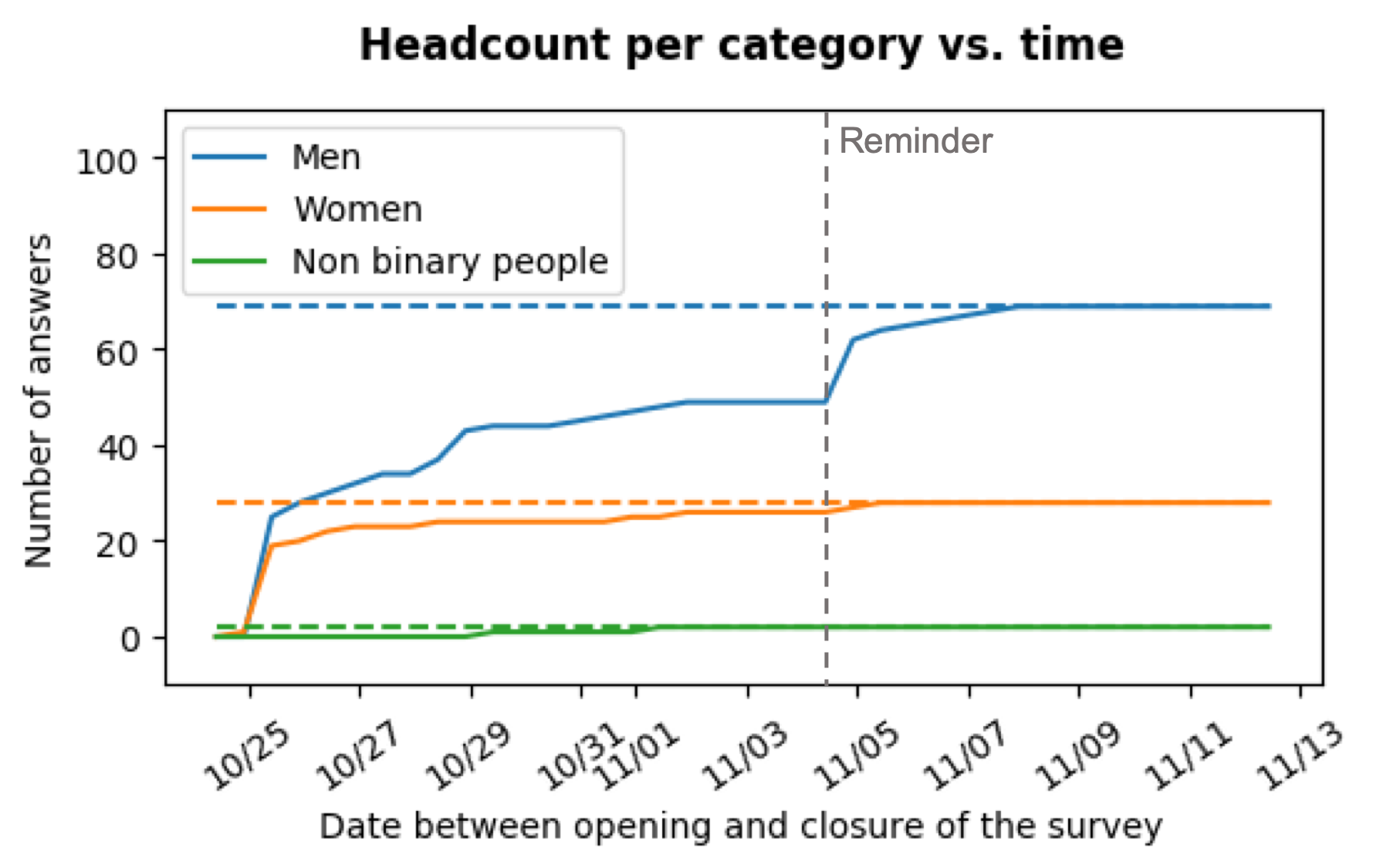} &
   \includegraphics[height=5cm]{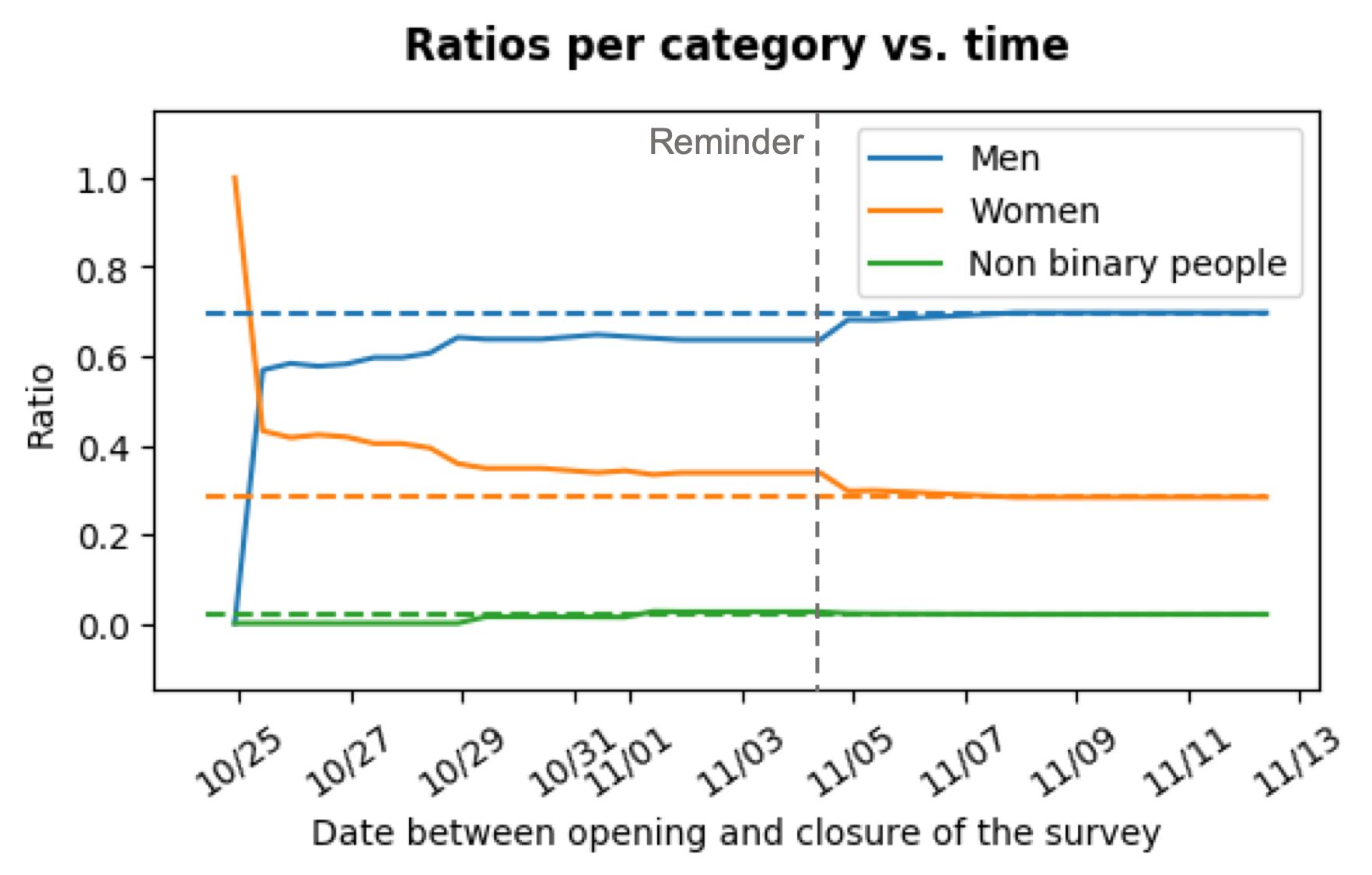}
   \end{tabular}
   \end{center}
   \caption 
   { \label{fig:RatiosvsTime} 
Evolution of the number of answers per gender with time (left) and of the proportion of each gender group with time (right). The vertical grey lines indicate the date of the reminder email. We can observe that women tend to answer faster than men to the survey, and the slight non-zero slopes on the right plot at the closure of the survey indicates an under-representation of male attendees to the survey and an over-representation of female attendees. The participation of non-binary attendees reaches a plateau well before the closure of the survey.}
   \end{figure*}

In order to analyze the interest of the conference participants to the survey and estimate the accuracy of the results per gender category, we monitored the evolution of the male and female respondents over three weeks during which the survey was open. The results are plotted in Fig.~\ref{fig:RatiosvsTime}, with  the number of participants per gender with time on the left and the fraction of each gender category with time on the right.

In Fig.~\ref{fig:RatiosvsTime}, left, we observe that the number of female and non-binary participants approach their final values very early on during the survey period (respectively within a few days and within a week of the opening date) compared to the male participation,  then show a near-flat slope. Conversely, the slope of the male participation near the end of the survey period (excluding the last 5 days when no new answer was received) is steeper. 
This different slope indicates that the survey is more likely missing answers from male participants than from female and non-binary participants and suggests a different level of interest to gender studies between these categories.

Similarly, assuming the three gender groups had a similar interest in the survey, they would have answered at the same rate and their fractions would have quickly stabilized around their final value in Fig.~\ref{fig:RatiosvsTime}, right, ie.  $69\%$ for men, $29\%$ for women, and $2\%$ for non-binary people. If we exclude the last 5 days when no answer was received, we see that this expected stabilization never occurs and that the final ratios evolve until the last answer. Later, this trend is even clearer when compared to the same analysis performed between the career level groups in Fig.~\ref{fig:InterestPosition}, right.

First, we conclude that we may have an over-representation of women and under-representation of men in this study compared to the total participants of the 2019 SOL conference. Second, it shows that women answered on average earlier than men, indicating a higher interest and involvement in the problems addressed in this survey. Finally, we notice that the reminder email sent on November 4th (vertical grey line of Fig.~\ref{fig:RatiosvsTime}) had a significant impact on the number of answers (+23), and mainly on the male participants (+20 answers).

Because of the low number of non-binary participants in the survey, the results for this sub-group suffer from large uncertainties and does not guaranty their anonymity in the next sections of the gender analysis. We thus chose to limit the rest of the analysis to the male and female genders only. We hope that conducting a similar study on a larger sample in the future will enable to report the responses of marginalized genders while preserving the anonymity of the people. 

\subsection{Distribution of career levels per gender}
\label{sec:GenderStatus}

Differences in the distribution of men and women between the main career status have been observed in many scientific communities, showing that women are more often graduate students than permanent researchers in comparison to male scientists (e.g. in the field of Adaptive optics, \citet{DOrgeville2014} or at the AGU Fall Meeting attenders, \citet{Ford2018}). Numerous factors have been studied in \citet{DOrgeville2014}, all leading to the conclusion that there is a leakage of women at each career step in the field of astronomy, ultimately leading to fewer women at higher positions. For the specific case of France, inequalities in access to the permanent positions is evident \citep{BerneHilaire2020}, with a success rate to permanent positions twice higher for men than for women. More generally, female scientists are less promoted and less funded than their male colleagues, generating a gap in the career distribution between the two genders \citep{Shen2013}. In this context, studying the career distribution per gender in our survey is of particular interest. 

\begin{figure}
   \begin{center}
   \includegraphics[height=4.5cm]{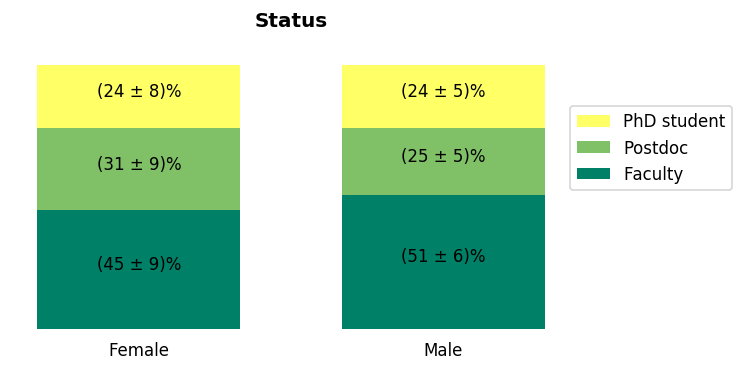}
   \end{center}
   \caption
   { \label{fig:GenderStatus} 
Career level distribution in both female and male gender groups. }
   \end{figure}

In Fig.~\ref{fig:GenderStatus}, we compare the professional positions of the male and female participants to the 2019 SOL conference survey. We see that the proportion of women with non-permanent positions is slightly higher than for men, with $31\% \pm 9\%$ of female postdocs versus $25\% \pm 5\%$ of male postdocs, and  $45\% \pm 9\%$ of female faculties versus $51\% \pm 6\%$ of male faculties. It thus seems that women have more difficulties to access permanent positions compared to male associates.  
However, the large uncertainties does not allow us to draw a solid conclusion here. This is a phenomenon commonly observed in other scientific communities that the fraction of women leaving academia accumulates throughout the classical research path, from graduation to full permanent position, in contrast to men associates (the \emph{leaky pipeline} phenomenon) \citep{DOrgeville2014, Ford2018}.
In Fig.~\ref{fig:GenderStatus}, we also observe that both gender groups have the same ratio of PhD students ($24\%$). 
It indicates that the general issue of women under-representation starts at the beginning of their career, with very few female students admitted in a PhD program in this field.


As discussed in Sec. \ref{sec:GenderImplication}, the extrapolation of the gender-based results to the whole Exoplanet Direct Imaging community may be limited by the likely under-representation of men in the survey answers. Yet, we show in Sec.~\ref{sec:Status} that the three career groups participated to the survey at similar rate, suggesting that the career level distributions are accurate and representative of this community.

\subsection{Exposure and visibility at the Spirit of Lyot 2019 conference}
\label{sec:BehaviorStatus}

   \begin{figure*}
   \begin{center}
   \begin{tabular}{cc}
   \includegraphics[height=4.5cm]{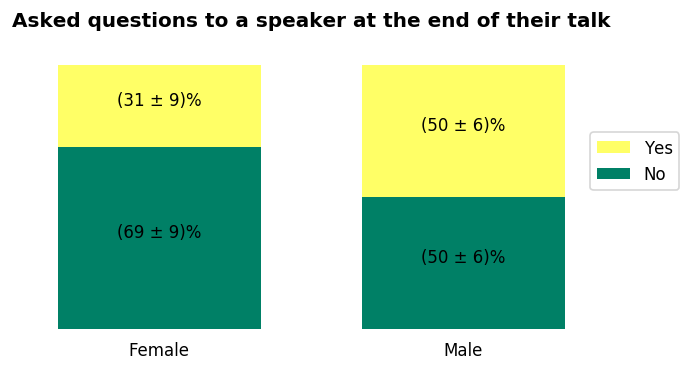} &
   \includegraphics[height=4.5cm]{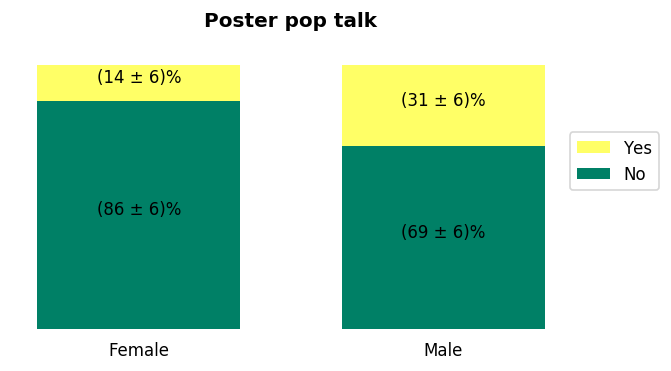}
   \end{tabular}
   \end{center}
   \caption
   { \label{fig:GenderTalk} 
Proportions of women and of men who asked questions at the end of a talk (left) and proportions of women and of men who asked for a poster pop talk (right) at the SOL conference.}
   \end{figure*} 

To analyse the exposure of each gender group at the 2019 SOL conference, several questions in the survey specifically asked if the participants requested a talk when they registered at the conference (question 7), and if they were actually attributed a talk by the SOC (question 8). 

From the answers, we derive that within the speakers, $67\%$ were men, $33\%$ were women, and none were non-binary people. The ratios for men and women are close to the fraction of female and male participants at the conference and indicate that the conference program was well representative of the binary attendees. However, we can regret that no non-binary person obtained an oral presentation. 

The results show that men and women equally asked for contributing talks ($78\% \pm 5\%$ for men vs. $76\% \pm 8\%$ for women) and that women were slightly more successful at obtaining one than men ($41\% \pm 6\%$ for men vs $48\% \pm 9\%$ for women), although the difference is not significant given the uncertainties. As a comparison, \cite{Ford2018} reports that at larger scales (AGU Fall Meetings from 2014 to 2016), women are given fewer opportunities than men to give oral presentations. As a reason for this observation, they explain that women are predominantly PhD students compared to men, and that students represent the least invited career group. In addition, they notice that conveners are mostly men, who seem less likely to give speaking opportunities to women.

The following results about self-confidence and self-promotion at the SOL conference are also derived from the specific questions inquired in the survey. Figure~\ref{fig:GenderTalk} (left) indicates that male attendees asked questions significantly more frequently than women following the talks ($50\%$ of the male participants vs $31\%$ of the female participants). This suggests that a larger fraction of women in this community are subject to self-censorship than men and did not consider the conference environment comfortable or benevolent enough to bring themselves forward. This behavior has also been observed and studied in other conferences such as the AAS meeting 223 \citep{Davenport2014} and the 2014 UK National Astronomy Meeting \citep{Pritchard2014} and on larger samples \citep{SchmidtDavenport2017}. In particular, the survey of the AAS meeting 223 \citep{Davenport2014} showed that the sessions chaired by women had a higher number of questions asked by women. 

Finally, a similar trend is observed for the poster-pop presenters: $31\%$ of the male participants vs $14\%$ of the female participants asked to advertise a poster on stage (Fig.~\ref{fig:GenderTalk}, right). 
Poster presentations were however open to all participants without selection by the SOC. This second point thus questions the appreciation of women for their own work and their confidence in front of an audience. 
We received explanations from 8 women for not volunteering to present their poster on stage, and sorted them in four categories: lack of information about the opportunity (50\%), lack of confidence (25\%), lack of interest in poster pops (12.5\%), and lack of time to prepare the poster pop (12.5\%). The similar analysis of the 13 explanations provided by male attendees also shows an important lack of information (31\%), but much more frequently a lack of interest (31\%) or a lack of time (23\%) for the exercise. Only one man suggested a lack of confidence, and one mentioned difficulties to have the poster presentation approved at the institutional level.

\subsection{Visibility and recognition by peers}
\label{sec:RecognitionGender}

   \begin{figure*}
   \begin{center}
   \begin{tabular}{cc}
   \includegraphics[height=4.5cm]{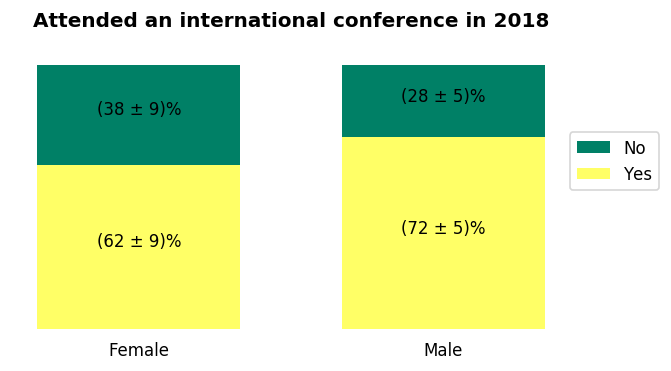} &
   \includegraphics[height=4.5cm]{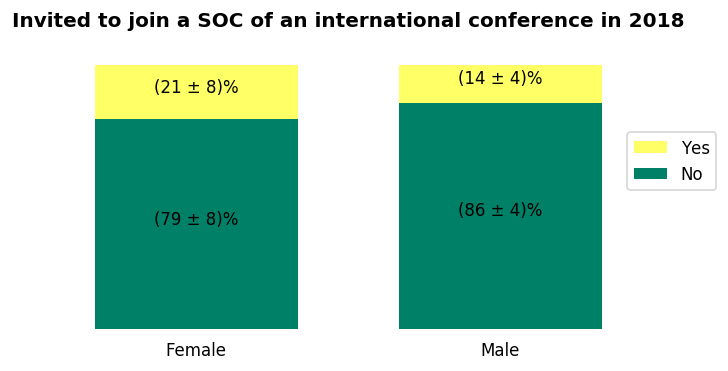}
   \end{tabular}
   \end{center}
   \caption
   { \label{fig:GenderConf} 
Proportions of women and of men who attended an international conference in 2018 (left) and  proportions of women and men having been invited in the SOC of an international conference in 2018  (right).}
   \end{figure*} 

In this section, we now extend the question of visibility and recognition to the daily professional environment instead of just 
the 2019 SOL conference. To address this topic, we focused on the general access to conferences, the participation to SOCs, and the inclusion in peer-reviewed publications as co-authors.

In terms of visibility and exposure to the community, $72\%$ of the male participants and $62\%$ of the female participants attended an international conference in 2018 (Fig.~\ref{fig:GenderConf}, left). This suggests that women may have higher difficulties to access conferences than men, although the sample was too small to guaranty the significance of this result. Given the importance of exposure at conferences to be recognized by peers, to advertise projects, and to promote one's career, the trend observed here in the Exoplanet Direct Imaging community should not be neglected and would need to be monitored and confirmed with a larger sample \citep{Prichard2019}.

In terms of recognition within the field, the survey showed that in 2018, slightly more women than men have been invited in the SOC of an international conference ($21\%$ of the female participants vs. $14\%$ of the male participants, see Fig.~\ref{fig:GenderConf} right). The uncertainties are however also too large to confirm this trend at a significant level. If confirmed on a larger sample, it may indicate an effort in this community towards a better representation of women at conferences. However, $50\%$ of the women having participated in a SOC in 2018 perceived that they were invited to fulfill a gender quota, and one of them specified that it was explicitly communicated to her. We note that the formulation of the question in the survey asking about the impression of filling a gender quota in a SOC may have been ambiguous about the time period considered and about the type of SOC (e.g. international conference vs. institutional committees, etc.), because two additional women answered affirmatively while indicating that they have not been invited in the SOC of an international conference in 2018. This ambiguity can be removed in future surveys by specifying that the question 17 refers to the same SOC as question 16. In any case, these numbers question the real motivation of increasing the representation of women in SOCs and on the recognition of their scientific expertise within this community.

   \begin{figure*}
   \begin{center}
   \begin{tabular}{cc}
   \includegraphics[height=4.5cm]{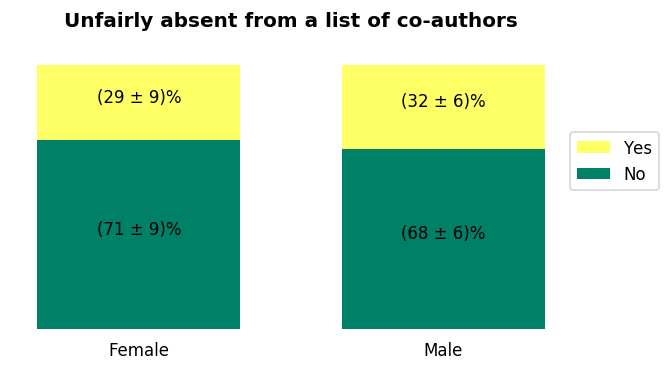} &
   \includegraphics[height=4.5cm]{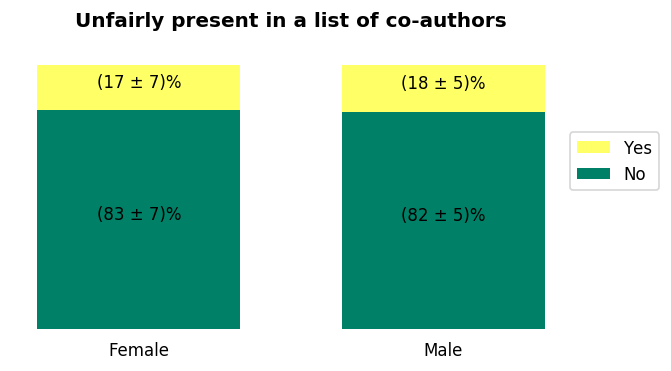}
   \end{tabular}
   \end{center}
   \caption 
   { \label{fig:GenderAuthor} 
Proportions of women and of men who felt being unfairly absent (left) and unfairly present (right) in a publication co-author list. }
   \end{figure*} 

Finally, the other aspect related to the recognition by peers probed in the survey concerned the inclusion in publications as co-author. Figure~\ref{fig:GenderAuthor} shows no significant difference between the fractions of men and of women who have sensed being unfairly excluded from a publication. We note that, independently of the gender considerations, a large fraction of people considers having been excluded from author lists ($\sim 30\%$ in total), which may indicate a larger issue. Likewise, we see no significant difference between the fractions of men and of women who felt unfairly present in the lists of co-authors ($\sim 18\%$). These two results indicate that there is no discrimination on gender in the inclusion in publication author lists within the Exoplanet Direct Imaging community. However, we note that, unlike the other questions of the survey, these two questions called on a subjective feeling and the responses were dependent on the sensitivity of the participant. This will be more discussed in the conclusion.

\subsection{Unprofessional behaviors per gender}
\label{sec:InapproriateBehaviorGender}

In this section, we analyzed the feedback to the survey questions probing the occurrence of and sensitivity to unprofessional behaviors in the Exoplanet Direct Imaging community. Similar to the previous section, some of these questions called on subjective interpretations of the respondents. 

   \begin{figure}
   \begin{center}
   \includegraphics[height=4.5cm]{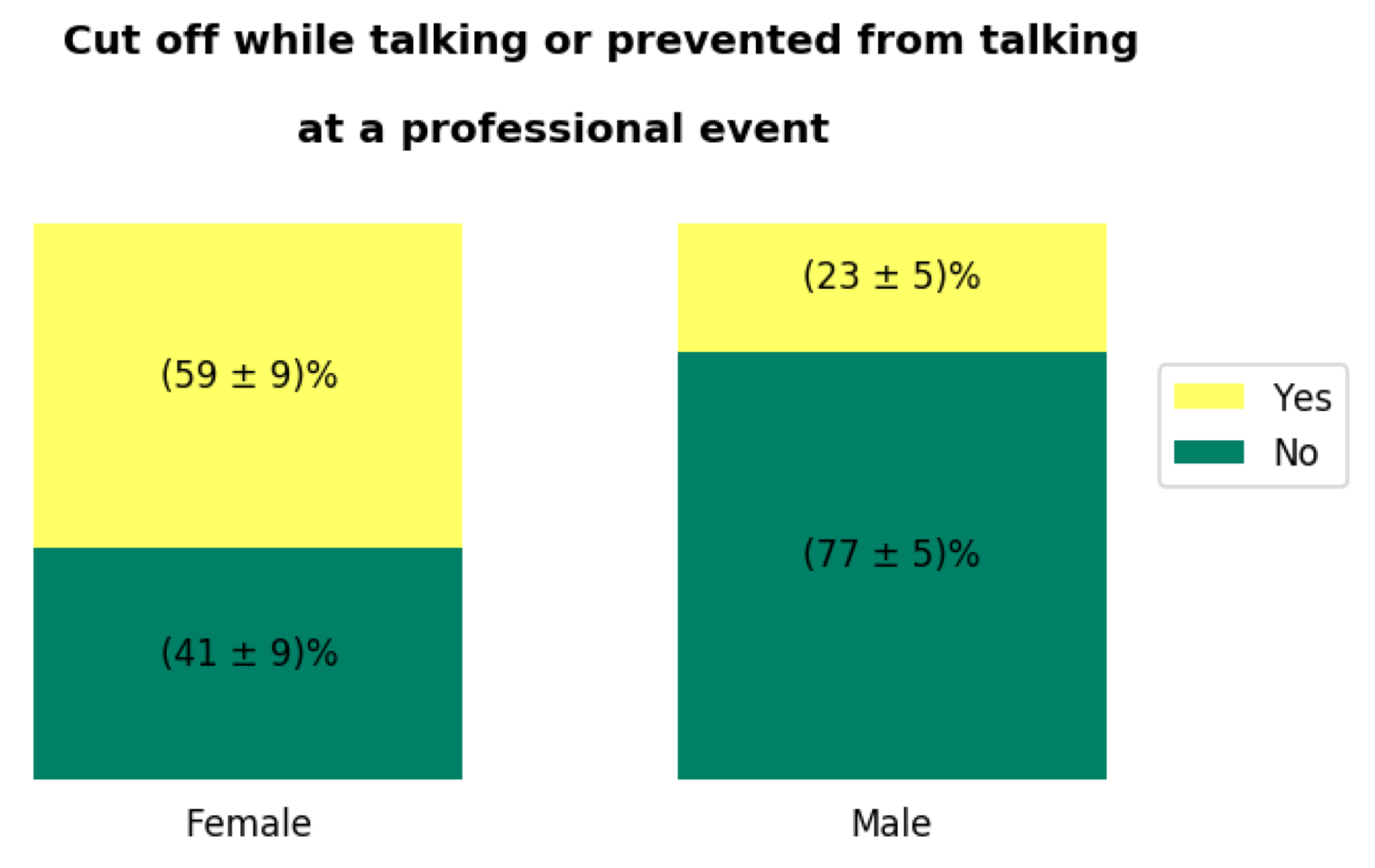}
   \end{center}
   \caption 
   { \label{fig:GenderCutOff} 
Proportions of women and of men who have already been interrupted or prevented from talking in a professional environment. A significant difference can be observed between the perception of both gender groups.}
   \end{figure} 

First of all, the survey inquired if the participants had already been interrupted while talking or prevented from talking at a professional event. Independently of the gender considerations, the total number of people who suffered from this behavior is 33\%. It reflects poor attitudes, indicating a lack of awareness in respecting boundaries among peers in the Exoplanet Direct Imaging community. In addition, the gender-based results revealed that $23\pm5\%$ of men versus $59\pm9\%$ of women consider they have faced this issue (Fig.~\ref{fig:GenderCutOff}). This shows a significant gender bias on this problem, with women being 2.6 times more frequently interrupted than men. This result is particularly appalling knowing that such conduct can impact one's self-confidence, leadership, and credibility, all indirectly linked to one's visibility, recognition, and thus career evolution. We recommend the community to be more respectful in this aspect and to be watchful against disrespectful interruptions, in particular of women.

   \begin{figure*}
   \begin{center}
   \begin{tabular}{cc}
   \includegraphics[height=5.2cm]{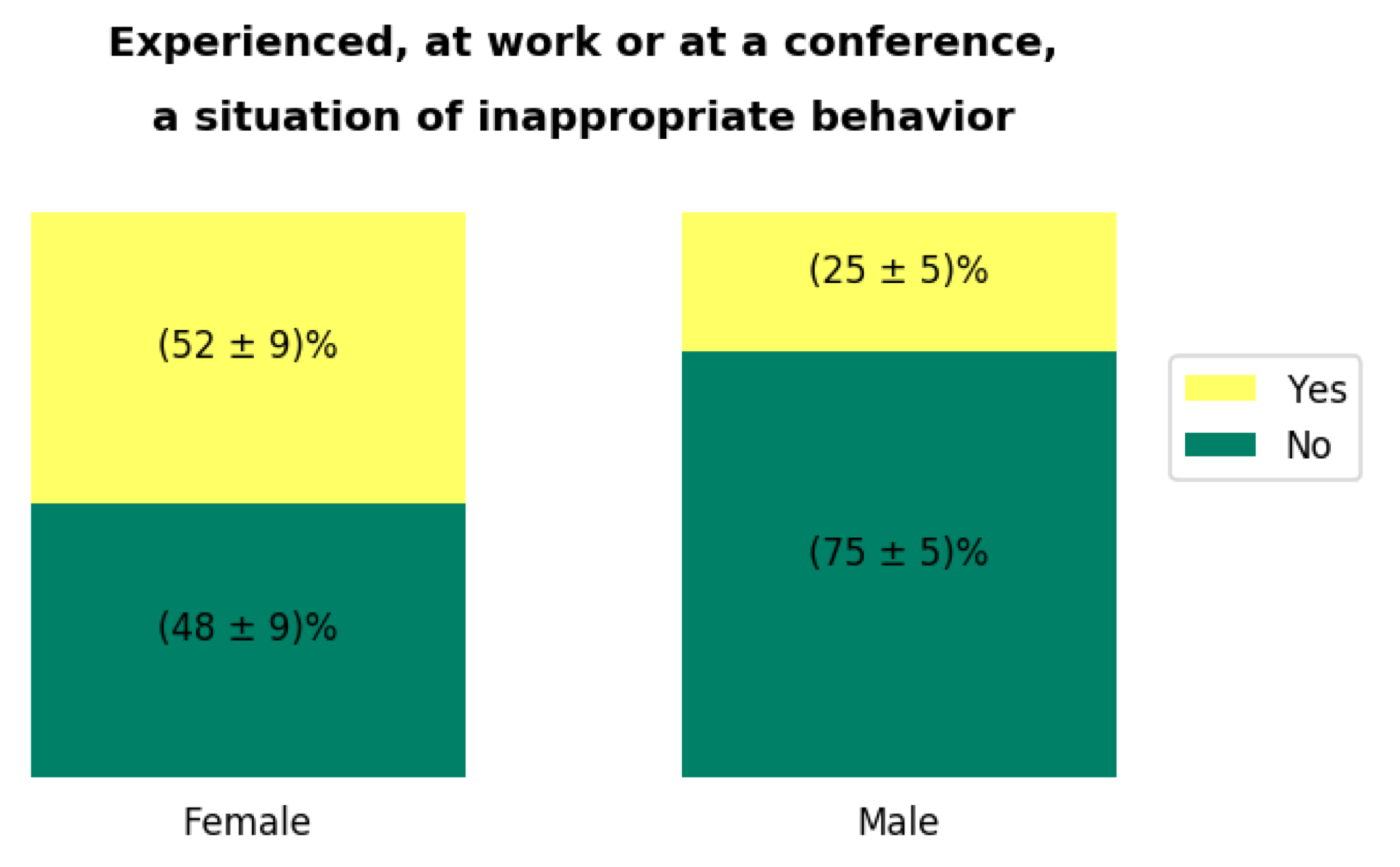} &
   \includegraphics[height=5.2cm]{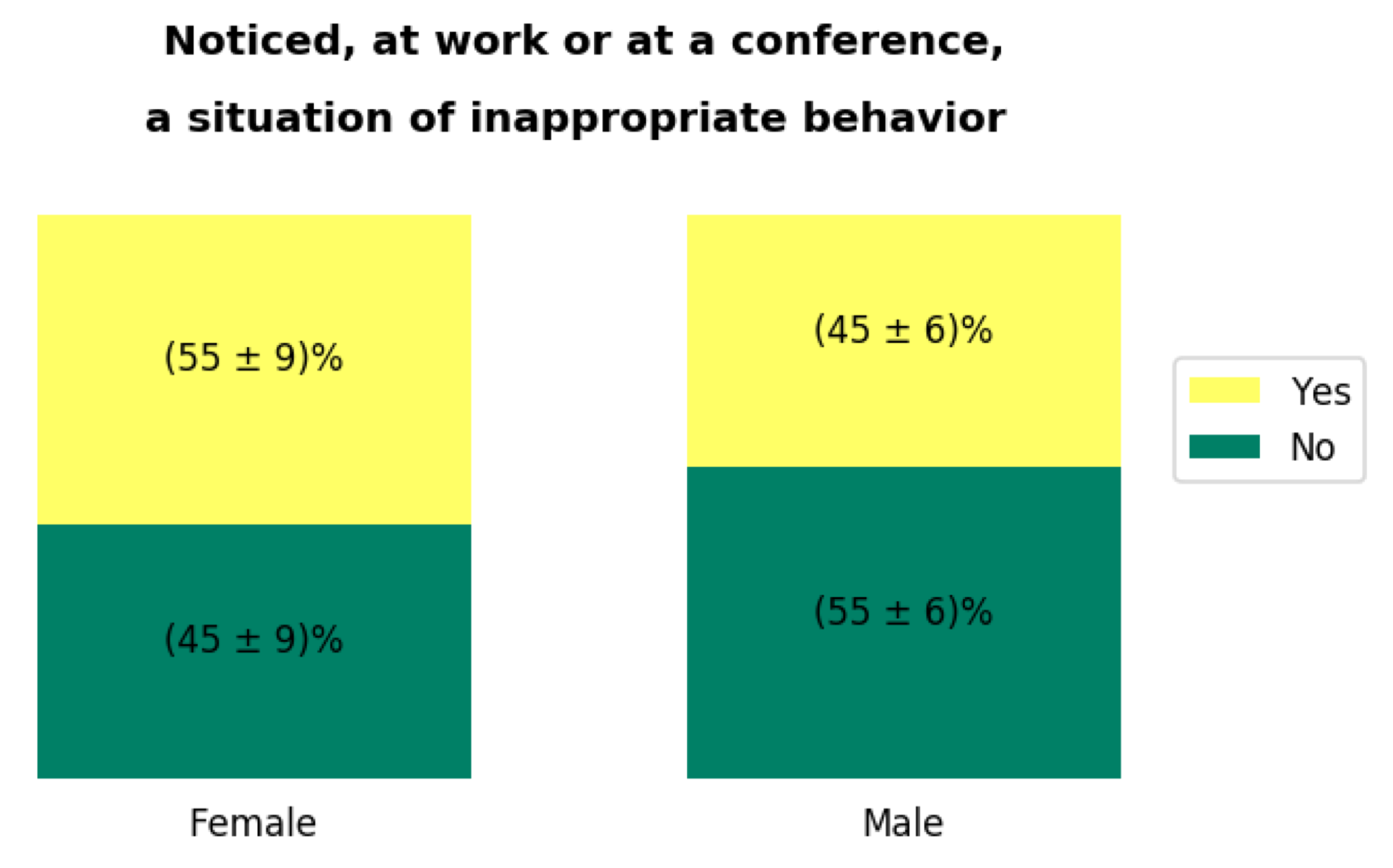} \\
   \includegraphics[height=5.2cm]{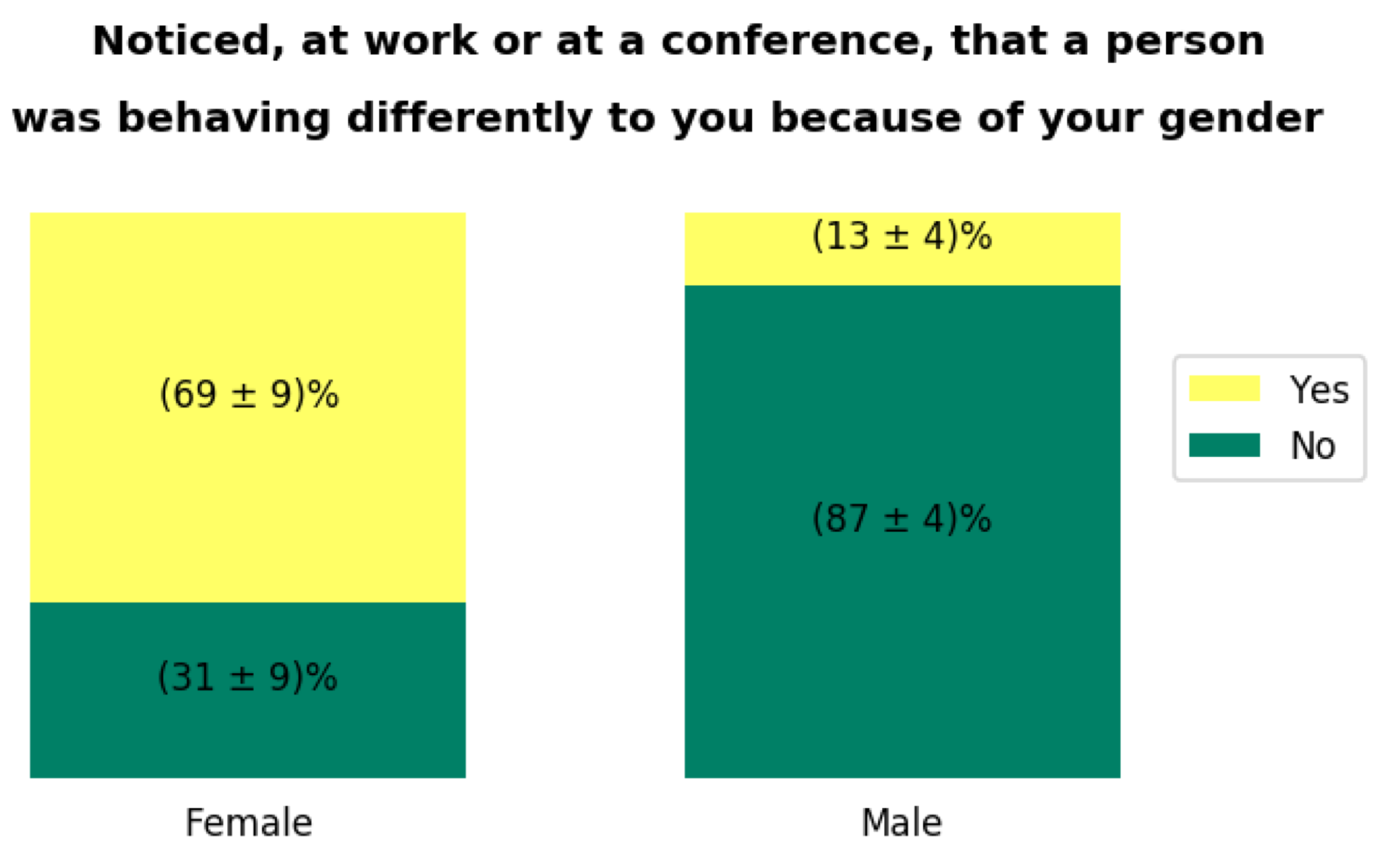} &
   \includegraphics[height=5.2cm]{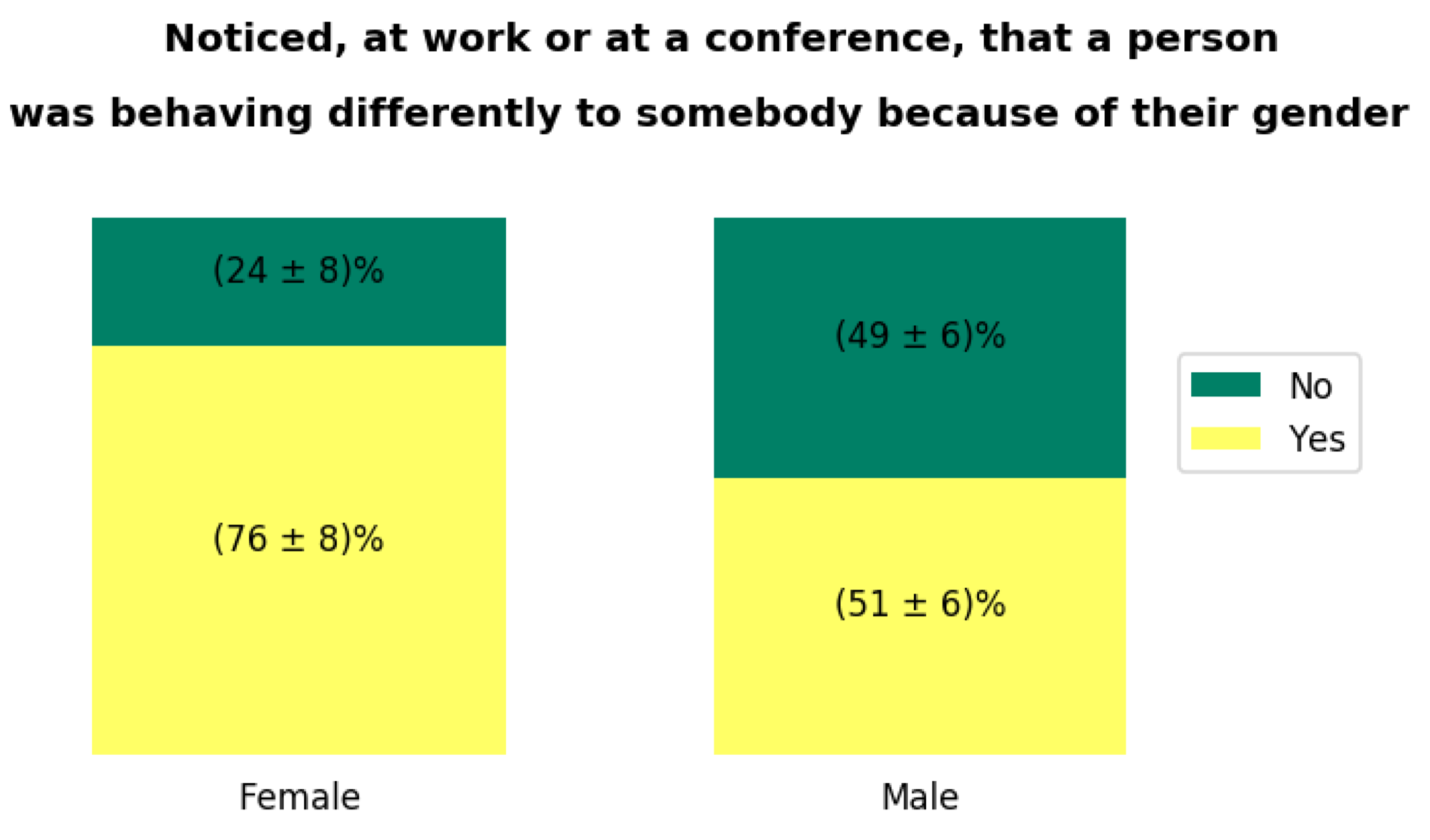}
   \end{tabular}
   \end{center}
   \caption
   { \label{fig:GenderInappropriate} 
Proportions of women and of men who experienced an inappropriate behavior (top left). Fractions of women and men who noticed a situation of inappropriate behavior (top right). Fractions of women and men who experienced a difference of behavior towards themselves due to their gender (bottom left).  Fractions of women and men who noticed a gender-based difference of behavior towards somebody (bottom right).}
   \end{figure*} 

Furthermore, $33\%$ of the participants have already been victims of an inappropriate behavior in this community. This rate is alarming as it shows a generalized and frequent behavior, revealing an unsafe work environment. It is also significantly unbalanced between men ($25\%$ of male victims) and women ($52\%$ of female victims), with 5.1 times more women experiencing inappropriate behaviors than men, as shown in Fig.~\ref{fig:GenderInappropriate} (top left). It implies that significant changes of conducts are urgent in the field of high-contrast imaging to make it a safe zone for everybody and for women in particular.

Complementary to these distressing results, we also observe that 48\% of the participants have already noticed a situation of inappropriate behavior, women slightly more frequently than men ($55\%$ vs. $45\%$, respectively, see Fig.~\ref{fig:GenderInappropriate}, top right). These values are somewhat encouraging as they indicate that people of both genders are aware of inappropriate behaviors and are capable of identifying such situations.

Finally, we obtain from the survey that a majority of the female participants (69\%) have experienced being treated differently because of their gender (see Fig.~\ref{fig:GenderInappropriate}, bottom left). In comparison, our data shows that this experience has happened to a negligible fraction of the male participants ($13\%$). Such a gender-based difference in behavior is not welcomed in a professional environment because it affects the credibility, leadership, and confidence of a person and influences his\textperiodcentered{}her\textperiodcentered{}their professional abilities. Thus we call for a significant change of conduct within the Exoplanet Direct Imaging community. Knowing that such gender-based changes of behavior may be performed unintentionally, we encourage the community to become further aware of its own prejudices, unintentional biases and be attentive and observing about one's own actions.

It is encouraging to note that 57\% of the participants have already noticed such gender-based differences of behavior. Although, Fig.~\ref{fig:GenderInappropriate} (bottom right) shows that women are significantly more vigilant about such behaviors than men ($76\%$ vs. $51\%$, respectively).

Overall, the results from this particular analysis draw a rather unsafe picture of the High-Contrast Imaging community. Combining the three types of unprofessional behaviors probed in the survey (questions 12, 14, 18), 54\% of the community have experienced at least one of these situations (gender-biased behavior: 29\%; oral interruption: 33\%; inappropriate behavior: 33\%). For women, the proportion of victims is 80\% (gender-biased behavior: 69\%; oral interruption: 59\%; inappropriate behavior: 52\%). Such events occur twice more frequently for women than for men, who are affected at 43\% by unprofessional behaviors.

These rates are alarmingly high and reveal an unsafe environment. Inappropriate behaving needs to be considered as a general issue and should be resolved urgently in our field. We observe that a large fraction of the community (around 50\%) generally notices inappropriate behaviors or gender-based differences. According to \citet{DOrgeville2014}, inappropriate behavior is one of the major problems causing women to leave academic research. The results described for the Exoplanet Direct Imaging community should thus be monitored and improved both to build a workplace safer for everybody and to push for a better gender representation.


\section{Results based on the career level}
\label{sec:Status} 
In this section, we analyze the results of the survey as a function of the professional position of the participants: Faculty member (or similar types of permanent position), postdoctoral researcher, or PhD student. About half of the participants were faculty members (47\%), and the other participants almost equally split between postdoctoral researchers (27\%) and PhD students (24\%), see Fig.~\ref{fig:General}. Two individuals in our data do not fit into these three predefined categories and answered "Other" in the career level options.

\subsection{Involvement in the survey}

\begin{figure*}
   \begin{center}
   \includegraphics[width=\hsize]{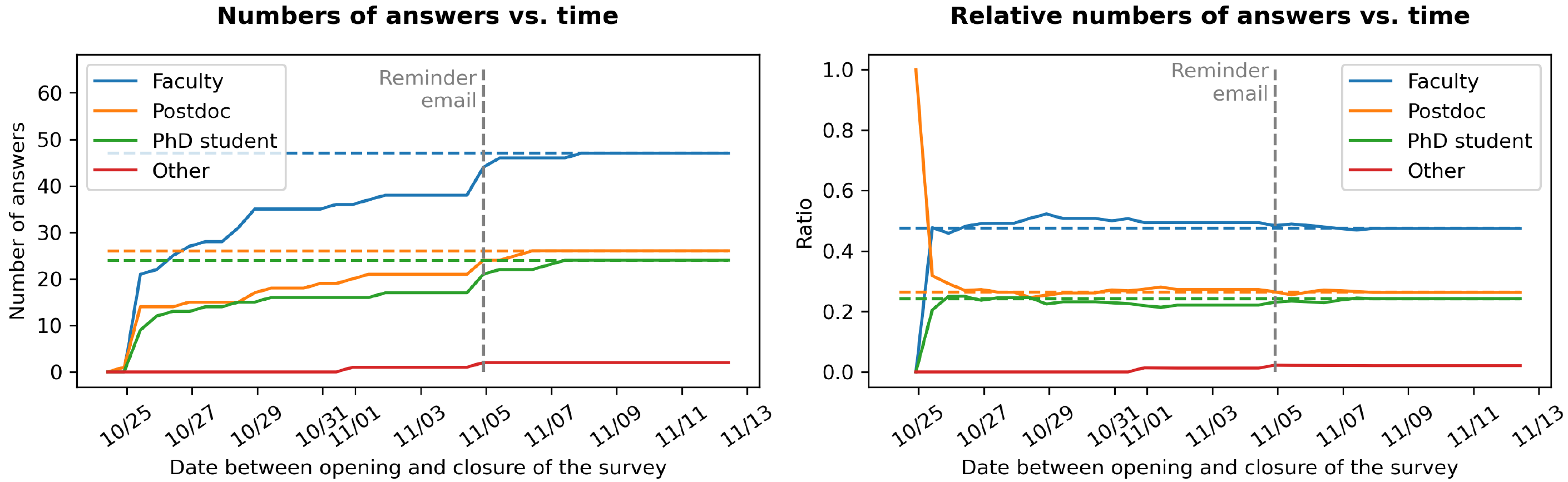}
   \end{center}
   \caption{ \label{fig:InterestPosition} 
Evolution of the number of answers to the survey with time per professional category, both in absolute values (left) and relative to the total number of answers per day (right). The ratios of answers reach a stable value for the three main professional  categories as early as the second day of the survey, unlike the gender balance rations. This shows a similar interest to the survey by all three major career categories.}
   \end{figure*}

Similar to the gender-based analysis, we analyzed the rate at which participants from the different career groups responded to the survey (Fig.~\ref{fig:InterestPosition}). First, we observe that the relative fractions between the three main career categories have all stabilized around their final values within the first two days of the survey period. Unlike the gender analysis, in which a bias against the survey was identified for the male group, the present analysis shows that all three career groups had a similar interest in the survey as their answers were received at the same rate. It also indicates that the survey accurately captured the actual ratios for each career group attending the Spirit of Lyot conference, despite the 53\% participation rate to the survey. 

\subsection{Gender balance per professional category}

   \begin{figure}
   \begin{center}
   \includegraphics[width=\hsize]{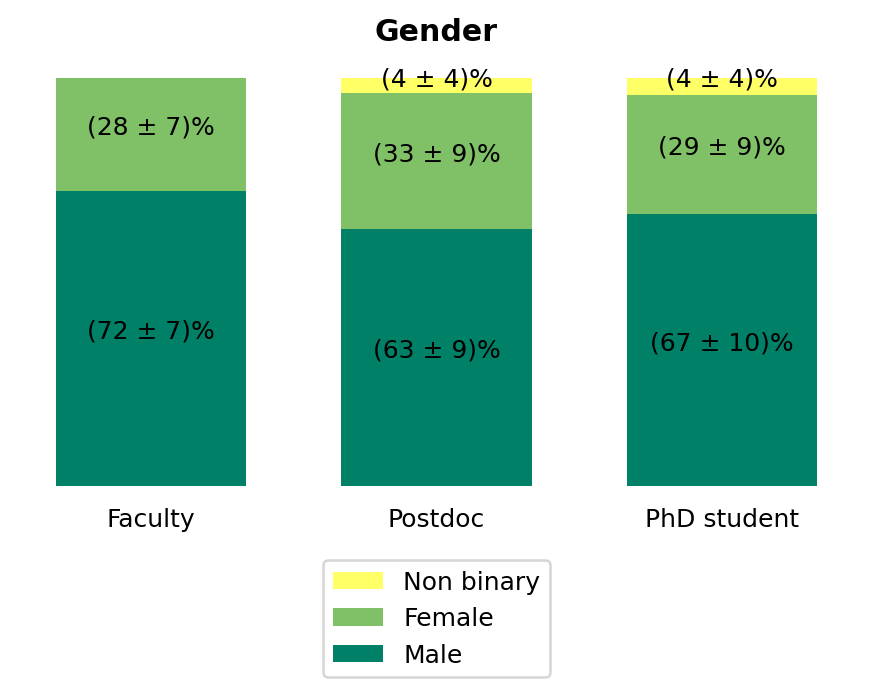}
   \end{center}
   \caption{ \label{fig:Position:Gender} 
Gender distribution per career group.}
   \end{figure} 

The current and the following sections focus only on the three major categories (Faculties, Postdocs, PhD students) and do not include the two "Other" entries as they are not statistically significant in this sample.

Figure~\ref{fig:Position:Gender} shows the gender balance in each professional category. It shows the same results as Fig.~\ref{fig:GenderStatus} but with a different perspective. 
Here, we see that gender imbalance exists in the three main career steps in the Exoplanet Direct Imaging community. 
Furthermore, the figure also aims at illustrating the so-called \emph{leaky pipeline} process in the community. At first glance, we observe a rather constant proportion of women at each career step (29\% of female PhD student, 33\% of female postdocs, 28\% of female faculties). This tentatively indicates that there is no obvious inequalities in the recruiting process within this community. However, the large error bars (7-9\%) are comparable to the ratio fluctuations between the career groups, which prevents us from drawing a firm conclusion for this field.

In addition, the low 29\% ratio of female PhD student is not encouraging to progress towards a better gender balance in the foreseeable future. Finally, we note that non-binary individuals equally spread in the younger PhD student and postdoc populations. This indicates that the Exoplanet Direct Imaging community is slowly starting to be more inclusive and diverse. Monitoring the evolution of these demographics in the next 5 years is necessary to assess whether these individuals are offered equal opportunities to obtain faculty positions.


\subsection{Expertise per professional level}

   \begin{figure}
   \begin{center}
   \includegraphics[width=\hsize]{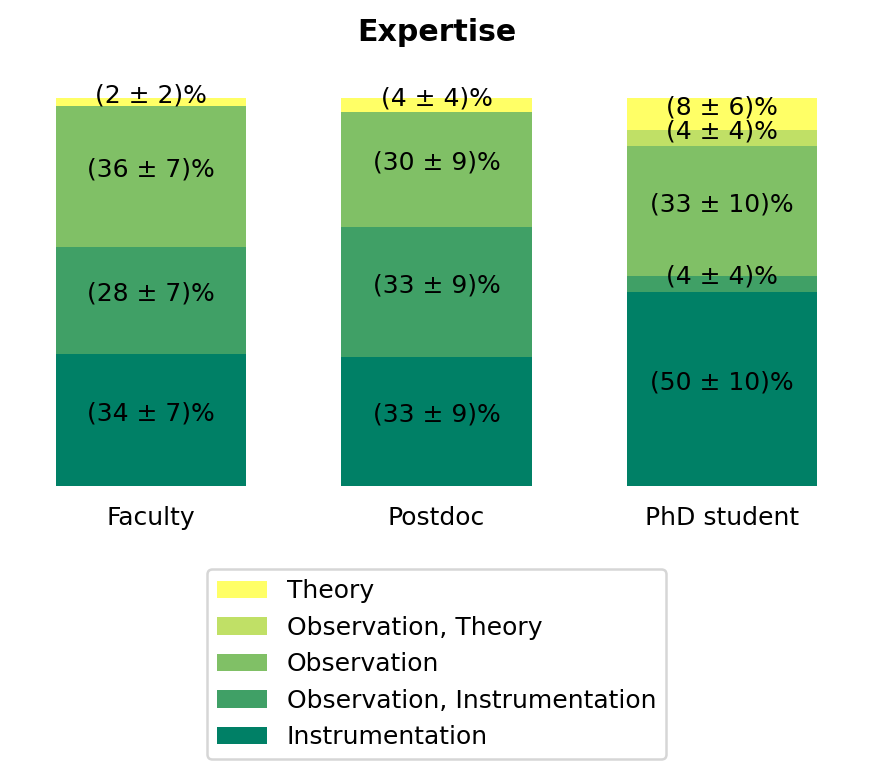}
   \end{center}
   \caption{ \label{fig:Position:Expertise} 
Expertise distribution per career group.}
   \end{figure} 

Figure~\ref{fig:Position:Expertise} shows the spread in expertise in each career group. We notice that the expertise is equally spread between "Instrumentation", "Observation", and "Observation and instrumentation" within the Faculty and Postdoc populations. PhD Students tend to have a single expertise (Instrumentation 50\%, Observations 33\%, Theory 8\%), but rarely a combination of several (8\% total). It reveals an (implicit) policy in this community to develop a strong expertise in a specific field at the PhD level and to diversify one's skills with additional expertise at the postdoc level. 

We notice that there are significantly more PhD students working in instrumentation (50\%) compared to the senior groups (30-36\%). It suggests that instrumentation is very attractive to the undergraduate students and is a good entry point for starting a career in astronomy. It also indicates that young doctors who specialized in instrumentation are more likely to broaden their expertise to observational skills than the ones with other expertise. This trend may be explained by several reasons: 1) there is a disinterest for instrumentation at the postdoc level, possibly explained by limited job opportunities in astronomical instrumentation or ample opportunities in private companies as compared to the other expertises, or 2) young instrumentalist doctors have more opportunities to develop observational skills, for instance after having commissioned an instrument or an instrument sub-system.

\subsection{Visibility and recognition by peers per career level}
   \begin{figure*}[!ht]
   \begin{center}
   \includegraphics[width=\hsize]{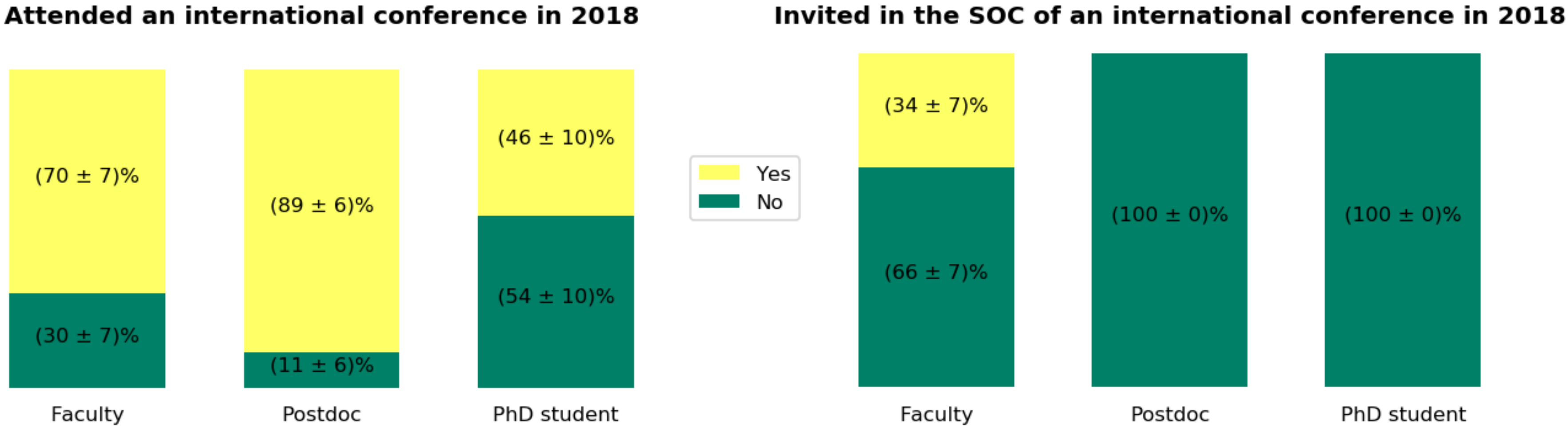}
   \end{center}
   \caption{ \label{fig:Position:Visibility1} 
Proportions of each career group who attended an international conferences in 2018 (left) and who were invited into the SOC of an international conference in 2018 (right).}
   \end{figure*} 

Figure~\ref{fig:Position:Visibility1} (left) shows the exposure given to each career category in international conferences. Figure~\ref{fig:Position:Visibility1} (right) presents the recognition level received by each career category within the community through invitations to join the SOC of an international conference. 

Our data shows that  postdocs receive the most exposure and visibility at international conferences, with 89\% of them who attended at least one conference in 2018. Counting the participation to the 2019 SOL conference, it shows that the large majority of postdocs attends an international conference at least once per year. This is a positive result, as the postdoctoral positions are based on short-term contracts, hence the postdocs are required to actively advertise their work and expand their network in search for a new position. Thus, this high visibility ratio demonstrates a healthy community which encourages and supports the postdocs in their career developments. 

However, only 46\% of the PhD student participants attended an international conference in 2018. 
We note that some of the PhD students attending the SOL conference may not have yet started their PhD program in 2018, so this percentage has to be taken carefully.
Nevertheless, combined with the low attendance of PhD students (24\%) as compared to the permanent researchers (47\%), it strengthens the previous analysis that this community does not promote its PhD students enough and prevents a fraction of them from attending conferences. This is yet another downside of the community, knowing that these events are critical for young researchers to promote their work and develop their professional network. This behavior is likely to make it more difficult for PhD students to find a postdoc position in the field.

In comparison, faculties receive ample amount of opportunities to regularly attend international conferences. In addition to attending the 2019 SOL conference, 70\% of the faculty participants attended an international conference in 2018. In addition to being the largest population of the SOL conference (47\% of the participants), this also strengthens the analysis that faculties in this field expose and promote themselves much more often than their PhD students.

Finally, Fig.~\ref{fig:Position:Visibility1}, right, shows that only permanent researchers have been invited to join the SOC of an international conference in 2018. Similarly, the SOC of the 2019 Spirit of Lyot conference was exclusively composed of faculty members. Although it can be argued that managing the scientific organization of conferences requires some level of professional experience, the exclusion of postdocs from SOCs demonstrates again a significant bias against young-career researchers in this community. We recommend the Exoplanet Imaging community to be more inclusive of young scientists in the future by supporting PhD students to attend international conferences and include postdocs in their scientific organization.

\subsection{Inappropriate behavior}

   \begin{figure*}
   \begin{center}
   \includegraphics[width=\hsize]{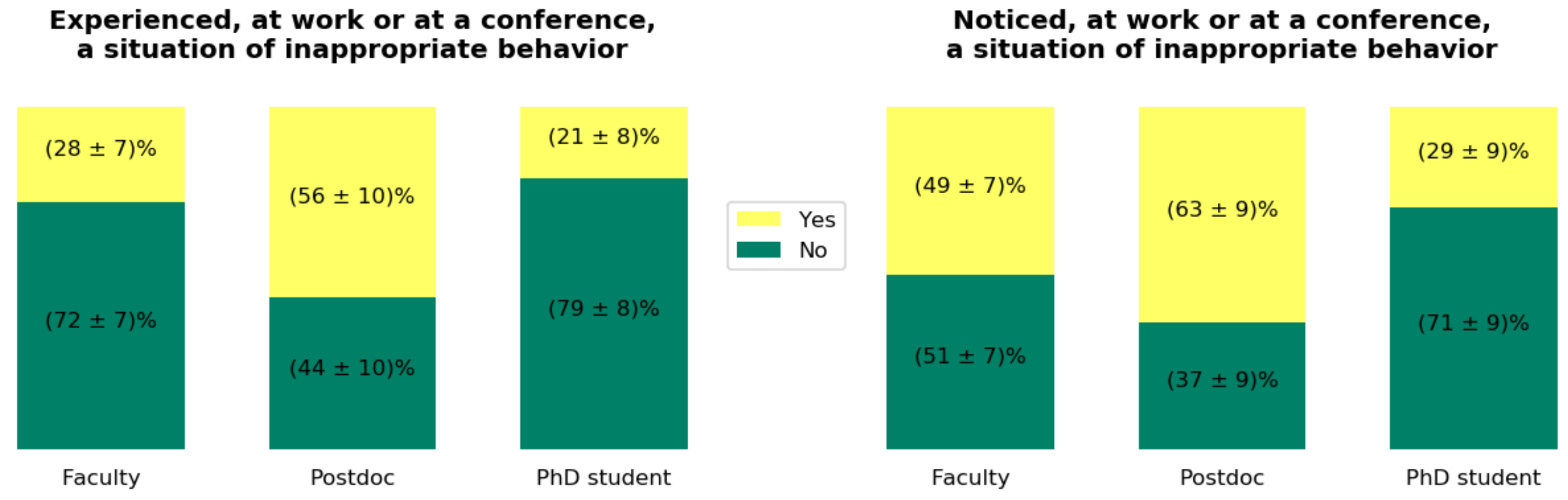}
   \end{center}
   \caption{ \label{fig:Position:inappropriate} 
Proportions of each career group who experienced (left) or noticed (right) situations of inappropriate behaviors.}
   \end{figure*} 

Figure~\ref{fig:Position:inappropriate} shows how the different professional categories are exposed to inappropriate behaviors. On the left, we show the fraction of each category that has experienced a situation of inappropriate behavior, and on the right the fraction that has noticed such behaviors.

In the left panel, we notice a significant difference between the professional categories in terms of experiencing an inappropriate behavior: the majority of postdoc participants have been victims of inappropriate behavior (56\%), about twice more than the permanent astronomers (28\%) and PhD students (21\%). Although all these ratios are high and concerning, this particularly high rate at the postdoc level is notably alarming given the vulnerability of postdocs to the professional insecurity, which is already a source of anxiety.

In the right panel, we notice that about half of the faculty community and a majority of postdocs have noticed a situation of inappropriate behavior at work or at a conference. On the one hand, this indicates that such behaviors happen relatively often in this community, which is a concern. On the other hand, it also indicates that the community is aware and attentive to such situations, with the postdoc community being the most vigilant, with 63\% of them having noticed inappropriate behaviors in the past. The PhD student community, however, seems relatively protected from exposure to the inappropriate behaviors, with 29\% of them having noticed such a situation.

It is particularly interesting to cross-correlate the occurrence of inappropriate behaviors between the gender and career groups, to identify more precisely the social groups which are most prone to such behaviors. In table~\ref{table:crossstudy}, we show the fractions of people having experienced inappropriate behaviors as a function of both their gender and career level. The non-binary and female postdoc group appear to be the most exposed to inappropriate behaviors, with 80\% of them who have  gone through such conduct. In comparison, male PhD students and male faculties are the most protected groups (19\% and 21\%, respectively). This is concerning and immediate steps should be taken to make this community safer, diverse and inclusive given how this well-known problem forces under-represented scientists (women and non-binary persons) to quit the field \citep{DOrgeville2014}. The fact that it happens predominantly at the postdoctoral stage, which is the most insecure in one's career, is particularly worrying as it discourages them to continue in the field and thus contributes to a vicious circle working against a representative gender-balanced community.

\begin{table}
\caption{Proportions of victims of inappropriate behaviors within each gender and career group. To preserve the anonymity of the non-binary participants despite their small sample, their answers are combined with those of women.
}
\label{table:crossstudy}
\centering
\begin{tabular}{lcc}
  \hline\hline
   & Female \& Non binary & Male \\
  \hline
  PhD students & 25 \%  & 19 \% \\
  Postdocs & 80 \% & 41 \% \\
  Faculty members & 46 \% & 21 \% \\
\end{tabular}
\end{table}

\subsection{Inclusion in publications}

\begin{figure}
   \begin{center}
   \includegraphics[height=4.5cm]{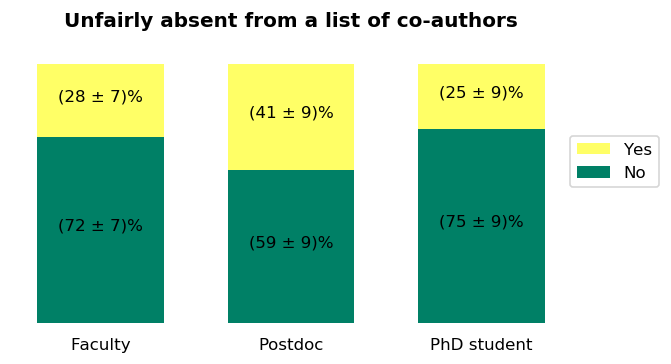}
   \end{center}
   \caption{ \label{fig:Position:authorlist} 
Proportions of each career groups who have already felt being unfairly absent from a publication author list.}
   \end{figure}

In Fig.~\ref{fig:Position:authorlist}, we show the fraction of people who mentioned being unfairly absent in the author list of a publication within the different professional categories. Here too we notice that the postdoc community most often declares that they have been unfairly left out from the publication authorship (41\% of them), compared to the faculty (28\% of them) and PhD student communities (25\% of them). This shared anxiety among the postdoc community may be enhanced by the insecurity of their short-term position and a need for a high publication rate to obtain a faculty position.

We do not report significant differences between these communities regarding the feeling of having been unfairly \emph{present} in author lists (17\% over all the participants regardless of their professional category).

\section{Discussion and conclusions}
\label{sec:Conclusions}

In this paper, we presented an overview of the community working on the direct imaging of exoplanetary systems based on a survey conducted at the Spirit of Lyot 2019 conference in Tokyo, Japan. The questions of the survey focused on several aspects about the general demographics, representation and self-confidence markers during this conference, the equity of exposure and recognition in the field through authorship, SOC invitation and access to conferences, and the occurrence of inappropriate behaviors at the workplace. The survey collected 100 answers, providing a 53\% participation rate from the conference attendees. In addition to the overall study, the results were analyzed within two main categories: as a function of the gender and of the career level of the participants.

From the global demographics analysis, we extract three main results: 
\begin{itemize}
    \item Women are under-represented in the community, with a 29\% representation rate.
    \item Young-career scientists (PhD students and postdocs) were under-represented at this conference, with a participation rate of 51\% compared to a representation in the field estimated between 66\% and 75\%.
    \item Significant representation disparities exist between the different countries present at the conference.
\end{itemize}

From the gender-based analysis, we gathered several key results. The principal positive results are the following: 
\begin{itemize}
    \item The proportions of PhD students, postdocs, and permanent researchers do not significantly vary between men and women. Owing to the large uncertainties, we cannot draw any conclusion about recruiting discrimination in the field. 
    \item At the Spirit of Lyot conference, the fractions of female contributing speakers (33\%) was comparable to the participation of women to the conference (29\%), showing an unbiased representation effort from the SOC. 
    \item Generally, in this field, a slightly higher fraction of women are invited in the SOC of international conferences compared to men  ($21\%$ of the female participants vs $14\%$ of the male participants), which shows an effort towards a better gender representation. However, a large fraction of these women (50\% of them) felt they were included in SOCs in order to fill a gender quota which questions the ability of the community to recognize them for their expertise.
\end{itemize}

However, several negative results are also reported: 
\begin{itemize}
    \item Women were more subject to self-censorship than men at the conference, which was deduced from two markers: they were significantly fewer to ask questions after the talks ($31\%$ of them vs $50\%$ of the male participants), and they were significantly fewer to advertise their poster on stage ($14\%$ of them vs $31\%$ of the male participants). The former behavior was also observed in other studies \citep{Davenport2014, Pritchard2014, SchmidtDavenport2017}.
    \item The most striking and alarming result is that 80\% of women reported having experienced unprofessional behaviors, ranging from gender-biased behaviors (69\% of them), to interruptions while they were speaking (59\% of them), to inappropriate behaviors (52\% of them). These rates are twice higher than for men (43\% of them experienced a type of unprofessional behavior). These behaviors contribute to creating an unsafe working environment and to undermining women's confidence, credibility, and leadership.
\end{itemize}
On an encouraging note, a majority of the participants, of all genders (although more often women), have already noticed such unwanted behaviors, which shows that the community is overall aware of these issues and can be in a position to intervene and stop such situations from happening.

From our career-based analysis, we mainly found that:
\begin{itemize}
    \item PhD students from this field are under-represented at conferences. This is seen both with their low participation to the Spirit of Lyot 2019 conference (24\%, 5 points lower than for instance the AGU meetings) and with the low fraction of them attending an international conference in 2018 (46\%). In comparison, faculties seem to more easily support themselves to conferences, with a 47\% fraction at the Spirit of Lyot conference and with 70\% of them attending an international conference in 2018. This may create difficulties for students in this community to find a postdoctoral position at the end of their thesis.
    \item Postdocs in this field are excluded from the SOC of international conferences. This strengthens the analysis that this community has a bias against young-career scientists.
    \item A majority of postdocs has already experienced inappropriate behaviors (56\%), significantly more than the PhD students and the faculties (around 25\% for each). In particular women and non-binary postdocs are predominantly victims of such situations (78\% of them). In comparison, 41\% of the male postdocs have encountered inappropriate situations. This is particularly alarming given that postdoc is the most insecure type of position: it may discourage early-career researchers especially female and non-binary scientists to continue working in this field.
\end{itemize}

This survey can be used as a reference point for future similar studies in order to monitor the evolution of the demographics and social behaviors in the field of direct imaging of exoplanetary systems. We also point out the systematic behaviors which need to be addressed so that the quality of the work environment can be improved. In the prospect of developing a safer and a well-represented community, we formulate a number of recommendations:
\begin{enumerate}
    \item Be proactive to prevent all sorts of professionally inappropriate behaviors, ranging from micro-aggression (interruptions, biased comments...) to serious aggression (harassment, intimidation,...). 
    \item Be more inclusive to under-represented genders, in particular at the PhD student recruiting level.
    \item In order to better promote young-career scientists, provide more support and encouragement to the PhD students to attend international conferences, and be more inclusive of postdocs when forming the SOC of conferences.
    \item Systematically implement a similar socio-demographic survey in the registration form of future conferences and workshops. 
\end{enumerate}
The later point has the capability of enabling the future studies to further monitor the evolution of gender representation and different biases over time, and allow to refine the solutions to counterbalance these issues. Furthermore, it would enable to increase the sample size and obtain more precise results. It would also provide awareness and more visibility to the experience of minorities including non-binary people. In particular, it was noted that many of the gender studies in science remain biased towards non-binary people and coming studies should follow the recommendations of \cite{Rasmussen2019} and and \cite{Strauss2020}. In order to help increasing the number of such studies, an improved version of the proposed survey is now available in open source on GitHub using this link: \href{https://github.com/lleboulleux/socio-demographic-community-survey-in-STEM}{https://github.com/lleboulleux/socio-demographic-community-survey-in-STEM}.
We encourage different communities to use it as a template to homogenize socio-demographic studies and allow long-term monitoring. We also invite the users of this survey to provide comments and feedbacks to improve its completeness. We are particularly interested in inputs and suggestions from researchers with a social science expertise. 

This study adds up to an increasing number of publications studying the community of researchers itself, instead of their research. While analyzing the results, we also listed improvements that could make this survey more precise on some aspects: 1) the questions about the inappropriate behaviors could be more specific about the different types of behaviors targeted in the study. These questions could even be repeated to probe the occurrence of a range of behaviors, from unwanted (e.g. gender-biased comments in a professional discussion) to inappropriate (e.g. sexual comments in a professional environment) to serious offences (e.g. aggression, harassment). 2) Some questions could be more specific on the period considered (e.g. questions 12--15, 17, 19, 20), which would remove ambiguities and make some career-based results more accurate. 3) Additional topics could be probed with this survey, such as the number of citations (see a very interesting study of gender-based differences on this topic by \citet{Caplar2017}), the occurrence of solicitations by scientific journals to review manuscripts \citep{LerbackHanson2017}, and the responsibility in large projects \citep{Piccialli2019}. More generally, social scientists are the experts for such studies and should be involved to guaranty that the survey questions are not phrased in a partial and suggestive way, but are expressed explicitly and neutrally. In addition, studies also acknowledge the experience of other under-represented groups in astronomy, such as LGBT \citep{Kay2009, Richey2019} or African American and Hispanic researchers \citep{Nota2009}.


\appendix

\section{Survey sent to all participants to the Spirit of Lyot 2019 conference}
\label{sec:Appendix}

The survey contained the following questions:

\noindent 1) Choose the option that best describes your gender (Female / Male / Non binary / Other / Prefer not to disclose)

\noindent 2) Choose the option that best describes your gender identity (Cisgender / Transgender / Other / Prefer not to disclose)

\noindent 3) What country do you currently live in?

\noindent 4) What kind of position do you occupy? (Intern / PhD student / Postdoc / Faculty)

\noindent 5) Did you attend an international conference in 2018?

\noindent 6) What is your expertise? (Instrumentation / Observation / Theory)

\noindent 7) At this conference (Lyot 2019): Did you ask for a talk?

\noindent 8) At this conference (Lyot 2019): Did you get a talk?

\noindent 9) At this conference (Lyot 2019): Did you ask for a poster pop talk?

\noindent 10) At this conference (Lyot 2019): If not, why?

\noindent 11) At this conference (Lyot 2019): Did you ask questions to a speaker at the end of their talk?

\noindent 12) HCI in general: Have you ever experienced, at work or at a conference, a situation of inappropriate behavior?

\noindent 13) HCI in general: Have you ever noticed, at work or at a conference, a situation of inappropriate behavior?

\noindent 14) HCI in general: Have you ever noticed, at work or at a conference, that a person was behaving differently to you because of your gender?

\noindent 15) HCI in general: Have you ever noticed, at work or at a conference, that a person was behaving differently to somebody because of their gender?

\noindent 16) HCI in general: Have you been invited to join a SOC of an international conference in 2018?

\noindent 17) HCI in general: Have you ever been invited to a SOC in order to fulfill a ratio of minority?

\noindent 18) HCI in general: Have you ever been cut off while talking or prevented from talking at a professional event?

\noindent 19) HCI in general: Have you ever been unfairly absent from a list of co-authors?

\noindent 20) HCI in general: Have you ever been unfairly present in a list of co-authors?

\noindent 21) Are there any comments you would like to share with us?

\acknowledgments

This work was co-authored by women and non-binary people having met at the Spirit of Lyot 2019 conference in Tokyo, Japan. It received full help and support from the LOC and the SOC of the conference and the participation rate of $53\%$ indicates an implication of the conference audience. We sincerely thank these three entities and particularly the SOC and LOC, who supported this initiative and helped advertise it among the conference attendees. The authors are sincerely grateful to the referee for their constructive comments, which improved the quality of the paper and  strengthened its conclusions. LL has received support of IRIS Origines et Conditions d’Apparition de la Vie (OCAV) of PSL Idex under the program Investissements d’Avenir with the reference ANR-10-IDEX-0001-02 PSL. GS would like to acknowledge the funding received from the European Union’s Horizon 2020 research and innovation programme under the Marie Sk\l{}odowska-Curie grant agreement No 798909. 

\bibliography{bib}{}
\bibliographystyle{aasjournal}

\end{document}